\documentclass[amstex,aps,preprint,nofootinbib,floatfix]{revtex4}%
\usepackage{amssymb}
\usepackage{graphicx}
\usepackage{amsmath}
\usepackage{bm}
\usepackage{amsfonts}%
\begin{document}
\title{Bohmian mechanics, the quantum-classical correspondence and the classical
limit: the case of the square billiard}
\author{A.\ Matzkin }
\affiliation{Laboratoire de Spectrom\'{e}trie physique (CNRS Unit\'{e} 5588),
Universit\'{e} Joseph-Fourier Grenoble-1, BP 87, 38402 Saint-Martin d'H\`{e}res, France}

\begin{abstract}
Square billiards are quantum systems complying with the dynamical quantum-classical correspondence. Hence an initially localized wavefunction launched along a classical periodic orbit evolves along that orbit, the spreading of the \emph{quantum} amplitude being controlled by the spread of the corresponding \emph{classical} statistical distribution. We investigate wavepacket dynamics and compute the corresponding de Broglie-Bohm trajectories in the quantum square billiard. We also determine the trajectories and statistical distribution dynamics for the equivalent classical billiard.  Individual Bohmian trajectories follow the streamlines of the probability flow and are generically non-classical. This can also hold even for short times, when the wavepacket is still localized along a classical trajectory. This generic feature of Bohmian trajectories is expected to hold in the classical limit. We further argue that in this context decoherence cannot constitute a viable solution in order to recover classicality.

\end{abstract}
\maketitle

\section{Introduction}

The de Broglie-Bohm (BB) theory of motion is generally regarded as the main
alternative to standard quantum mechanics (QM). The main achievement of the
theory in the non-relativistic domain is to deliver an interpretative
framework accounting for quantum phenomena in terms of point-like particles
guided by objectively existing waves along deterministic individual
trajectories \cite{holland}. Although the formalism does not give predictions
going beyond those of QM, it is often argued that BB should be favored because
of its interpretational advantages stemming from the ontological continuity
between the classical and the quantum domains. Thus, the so-called 'Bohmian'
trajectories followed by the quantum particle should be regarded as
objectively real as the trajectories of classical mechanics \cite{cushing96},
without the need to make a cut between the descriptions of reality at the
classical and the quantum levels \cite{bohm hiley85,home97}.

The aim of the present work is to investigate Bohmian trajectories in square
billiards and contrast them with the trajectories of the corresponding
classical system.\ A square billiard is the two-dimensional version of the
particle in a box problem, which was the example employed by Einstein in his
criticism of Bohm's rediscovery of de Broglie's pilot-wave \cite{einstein53}.
The interest of square billiards is that in terms of the quantum-classical
dynamical correspondence, the quantum mechanical propagator is constructed
from classical trajectories. Hence the quantum dynamics of a time-dependent
wavefunction is readily understood from the underlying classical dynamics --
each point of the wavefunction follows a classical trajectory. On the other
hand Bohmian trajectories are generically markedly different from their
classical counterpart: the Bohmian trajectories propagate by following the
probability flow, which results from the interference of several bits of the
wavefunction, each of which propagates by following a classical trajectory.
There is no criterion or limiting process (involving high energies,
macroscopic size, etc.) that will make the Bohmian trajectories resemble or
tend toward those of the classical billiard for a closed system.

Thus, although having BB trajectories irremediably different from the
classical ones in a closed system may not be a problem in itself, we will argue that when a quantum system displays the fingerprints of classical
motion, this creates
difficulties in view of the advantages traditionally attributed to the BB
interpretation.\ We will further contend that the way that is generally favoured
\cite{bohm hiley} in achieving the classical limit from Bohmian trajectories,
based on the decoherence resulting form the interaction of the system with its environment, suffers from a lack of consistency: we will question, in view of the quantum-classical correspondence, whether requiring localized and non-spreading wavefunctions is the correct way to define the classical limit for the BB interpretation.

We will proceed as follows.\ We will first give in Sec.\ 2 a brief account of
the classical square billiard, introducing the trajectories and the
propagation of classical statistical ensembles.\ Sec.\ 3 will deal with the quantum square
billiard, focusing on the propagation of initially localized wavepackets. We
will then (Sec.\ 4) give a brief overview of the de Broglie-Bohm theory and
display the Bohmian trajectories for the wavepackets previously shown in
Sec.\ 3. The results as well as their implications regarding the classical limit will be discussed in Sec.\ 5 and a summary with our
conclusions will be exposed in Sec.\ 6.

\section{The classical square billiard}

\subsection{Classical trajectories and periodic orbits}

A square billiard is a two dimensional box in the $(x,y)$ plane containing a
particle, which moves freely except for the specular bounces produced when it
hits one of the walls. Let $E=(p_{x}^{2}+p_{y}^{2})/2m$ be the total energy of
the particle and $L$ the length of one side of the square. Let $(x_{0},y_{0})$
denote the initial position of the particle. The classical trajectory followed
by the particle is readily obtained by integrating the equations of motion (it
is convenient to unfold the square box by propagating the trajectory in free
motion beyond the wall and then fold back the trajectory to the original
square \cite{robinett}). There are two types of orbits: either the particle
retraces the trajectory -- one obtains a periodic orbit (PO) --, or else the
trajectory covers entirely the billiard. Since the momentum is conserved, the
condition for a PO is that the trajectory appears closed in the $(x,y)$ plane,
which is possible if%
\begin{equation}
\frac{p_{x}}{p_{y}}=\frac{n_{y}}{n_{x}}\label{z2}%
\end{equation}
where $n_{x}$ and $n_{y}$ count the number of bounces off the $x$ and $y$ axes
respectively, and are therefore integers. A non periodic trajectory will be
obtained if $p_{x}/p_{y}$ is irrational.

Note that the PO condition only depends on the momenta: if a particle is
launched with the same momenta from two nearby initial positions, the two
periodic orbits will evolve in the same way, the PO's being deformed one relative
to the other, as in the example shown in Fig.\ 1(a). On the other hand if the second initial conditions also involve
a change in the momenta, the ensuing trajectory will not be periodic and will
deviate in time from the PO, as portrayed in Fig.\ 1(b). The period $T_{PO}$
of a periodic orbit is given by%
\begin{equation}
T_{PO}=L\frac{\sqrt{n_{x}^{2}+n_{y}^{2}}}{\sqrt{p_{x}^{2}+p_{y}^{2}}%
}.\label{z3}%
\end{equation}

\begin{figure}[tb]
\includegraphics[height=2.2in,width=4.7in]{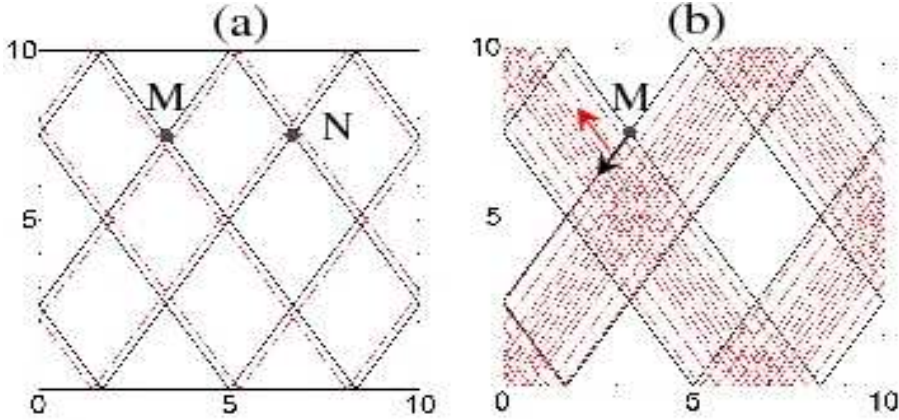}\caption{ Classical
trajectories in a square billiard with sides of length $L=10$ (arbitrary
units). (a): a classical periodic orbit (PO), going through the points labeled
$M$ and $N$ is shown in black. The grey dashed (online: red) line shows a trajectory
launched near $M$ with the \emph{same} momenta as the black PO; it is also a
PO. (b) Two trajectories are launched from $M$: the first one in black is the
PO shown in a). The second, in dashed grey (online: red) has slightly different
initial momenta, which is enough to render the trajectory non-periodic (the
red arrow shows the position of the trajectory slightly after $t\sim5T_{PO}$).
}%
\label{f1}%
\end{figure}

\subsection{Classical distributions dynamics}

The most general classical distribution $\rho(\mathbf{x},\mathbf{p},t)$ should
be considered in phase space. $\rho$ gives a density of particles having
positions $\mathbf{x}=(x,y)$ and momenta $\mathbf{p}=(p_{x},p_{y})$. The time
evolution of $\rho$ from an initial density $\rho_{0}(\mathbf{x},\mathbf{p})$
is governed by Liouville's theorem%
\begin{equation}
\frac{\partial\rho}{\partial t}=\left\{  H,\rho\right\}  ,\label{z5}%
\end{equation}
a statement of the conservation of the flow in phase-space, $d\rho/dt=0$;
$\left\{  {}, \right\}  $ denotes the Poisson bracket. Inside the billiard the
Hamiltonian $H=\mathbf{p}^{2}/2m$ is trivial. The bounces due to the wall
can be treated as above by considering first free motion for the distribution
and then appropriately folding it back inside the square. In terms of the
configuration space variables, Eq. (\ref{z5}) takes the form%
\begin{equation}
\frac{\partial\varrho}{\partial t}+\frac{1}{m}\nabla_{\mathbf{x}}\int
\rho(\mathbf{x},\mathbf{p},t)\mathbf{p}d\mathbf{p}=0\label{z7}%
\end{equation}
where
\begin{equation}
\varrho(\mathbf{x},t)\equiv\int\rho(\mathbf{x},\mathbf{p},t)d\mathbf{p}%
\end{equation}
is the configuration space density. Note that if the momentum is a
pre-assigned function of a position dependent momentum field $\mathbf{P}%
(\mathbf{x},t)$, the phase-space density takes the form $\rho(\mathbf{x}%
,\mathbf{p},t)=\varrho(\mathbf{x},t)\delta\left(  \mathbf{p-P}(\mathbf{x}%
,t)\right)  $ and Eq. (\ref{z7}) becomes%
\begin{equation}
\frac{\partial\varrho}{\partial t}+\frac{1}{m}\nabla_{\mathbf{x}}\left(
\varrho(\mathbf{x},t)\mathbf{P}(\mathbf{x},t)\right)  =0.\label{z9}%
\end{equation}
In classical mechanics, the field in configuration space is well-known to be given \cite{goldstein} in terms of the classical action $\mathcal{S}%
(\mathbf{x}_{0},\mathbf{x},t)$ via
\begin{equation}
\mathbf{P}(\mathbf{x},t)\equiv\nabla_{\mathbf{x}}\mathcal{S}(\mathbf{x}%
_{0},\mathbf{x},t)\text{,}\label{z8}%
\end{equation}
ensuring that the mechanical momentum is recovered. Note that $\mathbf{P}$ and
$\mathcal{S}$ are in general multivalued fields.

Classically, any normalized distribution can be envisaged.\ We will work with
initial distributions fairly well localized in configuration space. If each
point of the distribution has the same initial momentum obeying Eq.
(\ref{z2}), the propagation of the ensemble is straightforward -- the ensemble
moves along the family of neighboring periodic orbits as in Fig.\ 1(a).
However, anticipating on the analogy with the quantum mechanical square
billiard, we will choose initial distributions admitting a dispersion in the
momenta; the ensemble will then spread as it propagates.\ To be specific, let%
\begin{equation}
\rho_{0}(\mathbf{x},\mathbf{p})=\pi^{-2}\exp\left[  -\frac{(x-x_{0}){{}^{2}}%
}{2d{{}^{2}}}-2\Delta^{2}(p_{x}-p_{x_{0}})^{2}\right]  \exp\left[
-\frac{(y-y_{0}){{}^{2}}}{2d{{}^{2}}}-2\Delta^{2}(p_{y}-p_{y_{0}})^{2}\right]
\label{z10}%
\end{equation}
where $d$ and $\Delta$ are parameters that control the widths of the
Gaussians. By integrating over the momenta, we obtain%
\begin{equation}
\varrho_{0}(\mathbf{x})=\frac{\exp\left[  -\frac{(x-x_{0}){{}^{2}}}{2d{{}^{2}%
}}\right]  \exp\left[  -\frac{(y-y_{0}){{}^{2}}}{2d{{}^{2}}}\right]  }{2\pi
d^{2}}.\label{z12}%
\end{equation}
The usual properties of Gaussian distributions,%
\begin{align}
\left\langle x\right\rangle  &  =x_{0}\quad\left\langle x^{2}\right\rangle
=x_{0}^{2}+d^{2}\label{z13}\\
\left\langle p_{x}\right\rangle  &  =p_{x_{0}}\quad\left\langle p_{x}%
^{2}\right\rangle =p_{x_{0}}^{2}+1/4\Delta^{2}\label{z14}%
\end{align}
are verified (as well as the same properties for $y$). A solution of the
Liouville equation $\rho(\mathbf{x},\mathbf{p},t)$ follows by replacing
$\mathbf{x}\rightarrow\mathbf{x-p}t/m$ in Eq. (\ref{z10}); integrating over
the momenta yields%
\begin{equation}
\varrho(\mathbf{x},t)=\frac{\exp\left[  -\frac{2m^{2}\Delta^{2}}{t^{2}%
+4d^{2}m^{2}\Delta^{2}}\left(  (x_{0}-x+\frac{p_{x_{0}}t}{m})^{2}%
+(y_{0}-y+\frac{p_{y_{0}}t}{m})^{2}\right)  \right]  }{\frac{\pi t^{2}%
}{2\Delta^{2}m^{2}}+2\pi d^{2}}.\label{z15}%
\end{equation}
Hence in configuration space, the initially localized classical distribution
spreads in time, the spreading being controlled by the width of the initial
Gaussian $d$.\ We will further employ in this work only distributions
characterized by $\Delta=d$ (holding for the chosen units), so that the
product of the variances $V(x)V(p_{x}),$ readily obtained from Eqs.
(\ref{z13}) and (\ref{z14}) does not depend on $d$ or $\Delta$.

\begin{figure}[tb]
\includegraphics[height=4.in,width=5.5in]{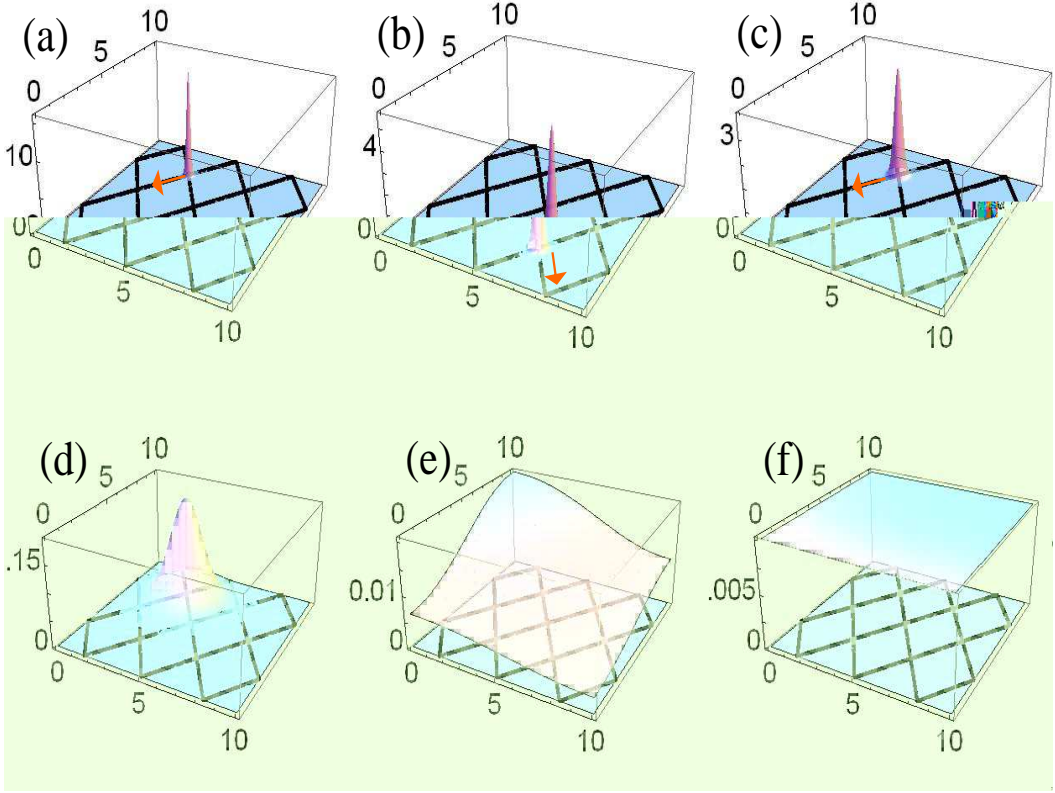}\caption{ Time evolution of
a \emph{classical distribution} in configuration space. (a) At $t=0$ the
classical distribution given by Eq. (\ref{z10}) is a Gaussian centered on $M$,
with $\mathbf{p}_{M}$ in the direction of the arrow, along the periodic orbit
shown in Fig. 1 (plotted in black). The height of the normalized distribution is given in
arbitrary units that are nevertheless the same in all the figures. The centre
of the distribution follows the periodic orbit. (b) gives a snapshot at
$t=3/4T_{PO}$ and (c) at $t=T_{PO} $ (first return at $M$). The initial
classical distribution spreads with increasing time; (d) shows the
distribution at $t=5T_{PO}$. For longer times, the distribution in the box
becomes a folded Gaussian: (e) shows the distribution at $t=25T_{PO}$ and (f)
at $t=100T_{PO}$, when the distribution is nearly uniform. }%
\label{f2}%
\end{figure}

An example is illustrated in Fig. 2: a distribution of the form (\ref{z10}) is
initially placed at $\mathbf{x}_{0}\equiv\mathbf{x}_{M}$ lying on the periodic
orbit shown in Fig. 1(a) with $\mathbf{p}_{0}\equiv\mathbf{p}_{M}$ in the
direction of the arrow along the PO.\ Fig.\ 2 shows snapshots taken at different times, as the
distribution spreads and becomes nearly uniform for $t\sim100T_{PO}$. Note
that due to the linearity of the Liouville equation, one can classically
envisage to take as the initial distribution the sum of two Gaussians
(\ref{z12}) localized at two different points of the billiard. The evolution
for the ensemble is then obtained by linear superposition of the evolution of
the two Gaussians, as illustrated in Fig.\ 3 for a distribution obtained by
superposing two ensembles initially localized on two points of the periodic
orbits $\mathbf{x}_{M}$ and $\mathbf{x}_{N}$.

\begin{figure}[tb]
\includegraphics[height=1.85in,width=5.5in]{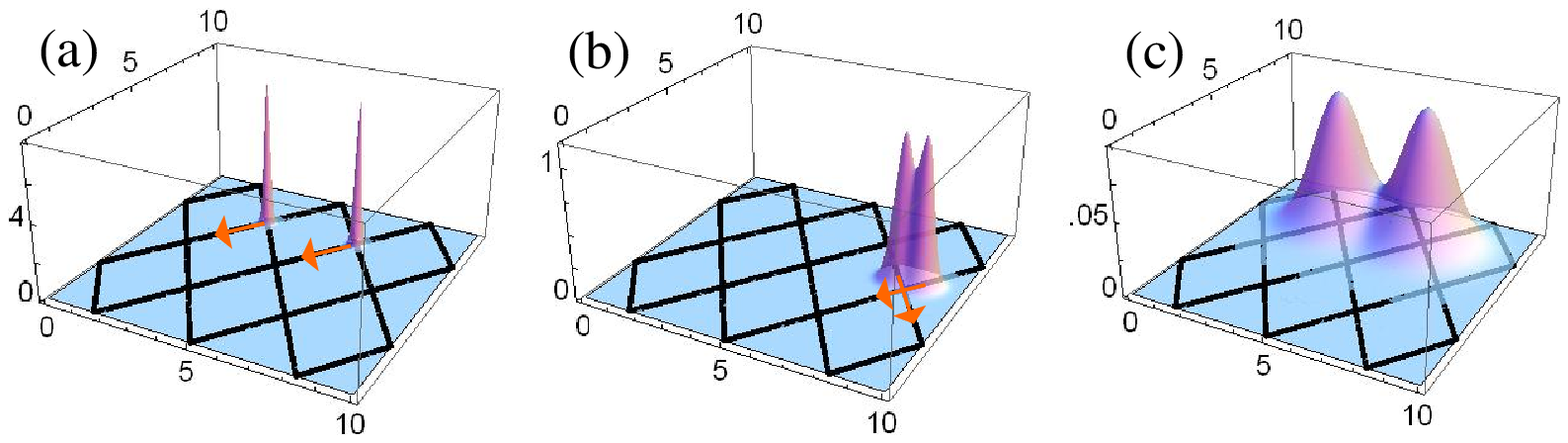}\caption{ Time evolution
of a \emph{classical distribution} composed of two initially localized
Gaussian components. (a) At $t=0$ the classical distribution is given by two
equal Gaussians centered at $M$ and $N$, with $\mathbf{p}_{M}$ and
$\mathbf{p}_{N}$ in the direction of the arrows, along the same periodic orbit
shown in Fig. 1. Each Gaussian follows the periodic orbit, spreading as the
time evolves. The two Gaussians must cross at several points before returning
to their respective initial points. (b) shows the distribution slightly before
the two Gaussians superpose when they cross at $t=(11/8)T_{PO}$. (c) shows the
situation at $t=5T_{PO}$. The initial Gaussians have sufficiently spread so
that their wings superpose. At longer times, one obtains the type of behaviour
shown in Fig. 2(f).}%
\label{f3}%
\end{figure}

\section{The quantum square billiard}

\subsection{Eigenstates and propagator}

The eigenstates in configuration space and eigenvalues of the quantum billiard
are readily obtained from those of the infinite well problem,%
\begin{align}
\psi_{n_{x},n_{y}}(x,y)  &  =\frac{2}{L}\sin\frac{n_{x}\pi x}{L}\sin
\frac{n_{y}\pi y}{L},\\
E(n_{x},n_{y})  &  =\frac{\hslash^{2}\pi^{2}}{2mL^{2}}(n_{x}^{2}+n_{y}^{2}).
\end{align}
The propagator -- the configuration space representation of the time evolution
operator -- is that of the free particle with an appropriate folding into the
original square. The free particle propagator takes the well known form%
\begin{equation}
K(\mathbf{x}_{0},\mathbf{x},t)=\frac{m}{2\pi\hslash t}\exp\left(  \frac
{im}{2\hslash t}\left[  (x_{0}-x)^{2}+(y_{0}-y)^{2}\right]  -i\frac{\pi}%
{2}\right)  ,\label{z20}%
\end{equation}
which in the square billiard is exact for short times (no bounces on the
walls). We will omit to give explicitly the additional terms accounting for
the bounces that must be added to Eq. (\ref{z20}) (see eg Ch.\ 6 of
\cite{kleinert} for the full expression). Instead it will be more convenient for interpretational purposes
to employ the semiclassical form of $K$.\ Recall that
for free or quadratic potentials, the semiclassical approximation to the
propagator is exact: the semiclassical propagator is given by \cite{schulman}%
\begin{equation}
K(\mathbf{x}_{0},\mathbf{x},t)=\sum_{k}\frac{1}{2i\pi\hslash}\left\vert
\det\frac{\partial^{2}\mathcal{S}_{k}}{\partial\mathbf{x}\partial
\mathbf{x}_{0}}\right\vert \exp\left(  i\mathcal{S}_{k}(\mathbf{x}%
_{0},\mathbf{x},t)/\hslash+i\phi_{k}\right)  ,\label{z21}%
\end{equation}
where the sum runs on all the classical trajectories $k$ connecting
$\mathbf{x}_{0}\mathcal{\ }$to $\mathbf{x}$ in the time $t$. $\mathcal{S}_{k}$
is the classical action for the $k$th trajectory and the determinant is the inverse of the
Jacobi field familiar from the classical calculus of variations, reflecting the local density of the paths. $\phi_{k}$ is
a phase that takes into account the bounces on the hard wall.

\subsection{Quantum dynamics}

We will take for the initial wavefunction the localized Gaussian%
\begin{equation}
\psi_{0}(\mathbf{x})=\frac{1}{d\sqrt{2\pi}}\exp\frac{-(x-x_{0})^{2}%
-(y-y_{0})^{2}}{4d^{2}}\exp\frac{i}{\hslash}\left(  xp_{x_{0}}+yp_{y_{0}%
}\right)  .\label{21}%
\end{equation}
$p_{x_{0}}$ and $p_{y_{0}}$ can be taken as parameters, though their physical
meaning is revealed by taking the Fourier transform or computing the averages%
\begin{align}
\left\langle \hat{X}\right\rangle _{\psi_{0}}  &  =x_{0}\quad\left\langle
\hat{X}^{2}\right\rangle _{\psi_{0}}=x_{0}^{2}+d^{2}\label{z22}\\
\left\langle \hat{P}_{x}\right\rangle _{\psi_{0}}  &  =p_{x_{0}}%
\quad\left\langle \hat{P}_{x}^{2}\right\rangle _{\psi_{0}}=p_{x_{0}}%
^{2}+\hslash^{2}/4d^{2},\label{z23}%
\end{align}
which unsurprisingly are the same as the classical ones if one puts
$\Delta\sim d/\hslash$.

\begin{figure}[tb]
\includegraphics[height=3.9in,width=5.5in]{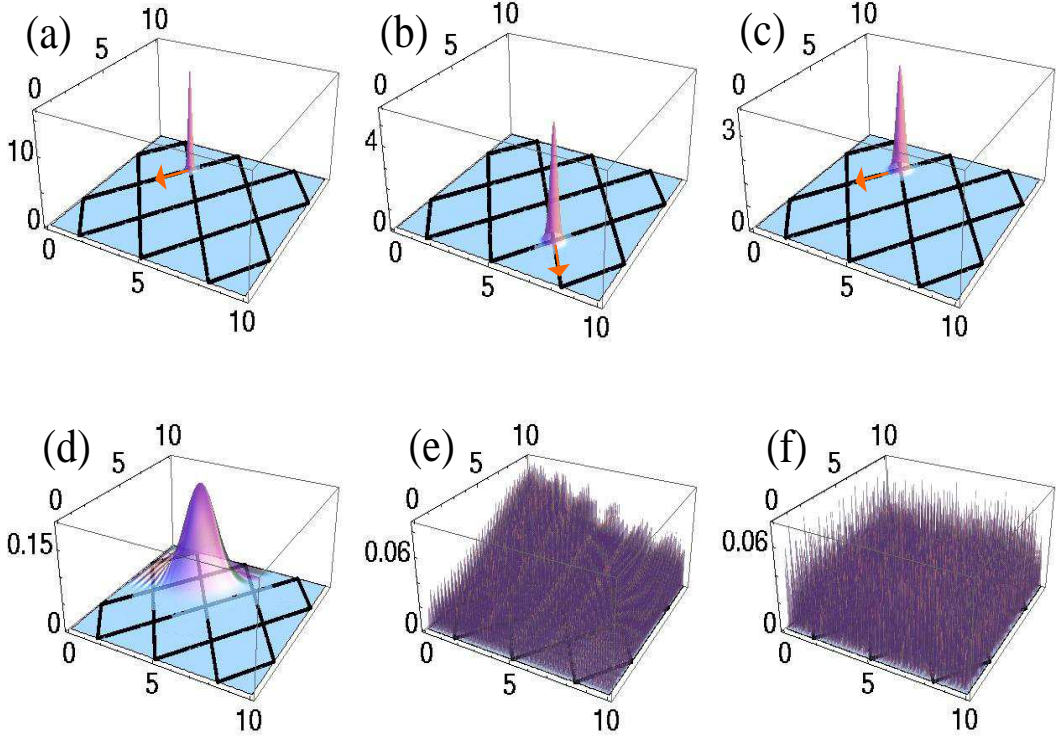}\caption{ Time evolution of
the \emph{quantum probability density} in configuration space. (a) At $t=0$
the initial wavefunction is given by Eq. (\ref{21}) with $\mathbf{x}_{0}%
\equiv\mathbf{x}_{M}$ and $\mathbf{p}_{0}\equiv\mathbf{p}_{M}$. The parameters
are chosen to match exactly those of the classical distribution illustrated in
Fig. 2 (hence we take $\hbar=1$). With these parameters, the wavefunction
leaves the region around $M$ in the direction of the arrow and propagates
along the classical periodic orbit. (b), (c) and (d) give snapshots of
$\left\vert \psi\right\vert ^{2}$ at $t=3/_{4}T_{PO}$, $t=T_{PO}$,
and$t=5T_{PO}$ respectively. The evolution and spreading of the quantum
probability density is nearly identical to that of the classical counterpart
in Fig. 2. (e) and (f) shows the quantum probability distribution at
$t=25T_{PO}$ and $t=100T_{PO}$ respectively. For longer times, interference
resulting from the reflection of the distribution as it spreads creates a very
high density of peaks, though as in the other cases, the smoothed quantum distribution corresponds
to the classical one. }%
\label{f4}%
\end{figure}

The time evolved wavefunction,
\begin{equation}
\psi(\mathbf{x},t)=\int d\mathbf{x}^{\prime}K(\mathbf{x}^{\prime}%
,\mathbf{x},t)\psi_{0}(\mathbf{x}^{\prime})\label{z24}%
\end{equation}
is readily computed by employing Eq. (\ref{z20}), giving the probability
density%
\begin{equation}
\left\vert \psi_{ST}(\mathbf{x},t)\right\vert ^{2}=\frac{\exp\left[
-\frac{2m^{2}}{\hslash^{2}t^{2}/d^{2}+4d^{2}m^{2}}\left(  (x_{0}%
-x+\frac{p_{x_{0}}t}{m})^{2}+(y_{0}-y+\frac{p_{y_{0}}t}{m})^{2}\right)
\right]  }{\frac{\pi\hslash^{2}t^{2}}{2d^{2}m^{2}}+2\pi d^{2}}.\label{z25}%
\end{equation}
This expression is exact for the free particle, but is only valid for
\emph{very short times} in the square billiard. Note nevertheless that the
probability density propagates exactly like the classical distribution
(\ref{z15}).\ The additional terms due to the bounces that need to be added to
Eq. (\ref{z21}) can produce interferences at longer times, as is apparent by
using the expression (\ref{z21}) of the propagator. Indeed, if there are
several points $\mathbf{x}^{\prime}$ such that $\psi_{0}(\mathbf{x}^{\prime})
$ is non-vanishing that are propagated to the same $\mathbf{x}$ in the time
$t$, different trajectories will contribute to $K$ in Eq. (\ref{z24}), leading
to interferences. This means that each point $\mathbf{x}^{\prime}$ of the
initial wavefunction is carried by a classical trajectory to the final point
$\mathbf{x},$ interferences happening when several classical trajectories each
carrying a part of the propagating wavefunction arrive simultaneously at
$\mathbf{x}$.

Summarizing, we can say that the quantum propagation of the initial
wavefunction is exactly like the propagation of an analog classical
distribution, except for the provision of the wavefunction superposition
(whereas in the classical case the superposition concerns the positive valued
distribution themselves). Examples are given in Figs. 4 and 5.\ In Fig. 4 the
initial wavefunction is of the form (\ref{21}) with $\mathbf{x}_{0}%
\equiv\mathbf{x}_{M}$ and $\mathbf{p}_{0}\equiv\mathbf{p}_{M}$ in perfect
correspondence with the classical distribution lying on the periodic orbit
shown in Fig.\ 2. Fig.\ 5 combines in the initial wavefunction two Gaussians
at $\mathbf{x}_{M}$ and $\mathbf{x}_{N}$ with respective parameters
$\mathbf{p}_{M}$ and $\mathbf{p}_{N}$ matching those of the classical
distribution pictured in Fig.\ 3. We take as the initial wavefunction%
\begin{equation}
\psi(\mathbf{x},t=0)=\frac{1}{\sqrt{2}}\left(  \psi_{M}(\mathbf{x})-\psi
_{N}(\mathbf{x})\right) \label{4}%
\end{equation}
which is one choice among many other possibilities leading to an initial
quantum density matrix whose diagonal elements in the position representation match the initial classical distribution of Fig.\ 3.

\begin{figure}[tb]
\includegraphics[height=1.85in,width=5.5in]{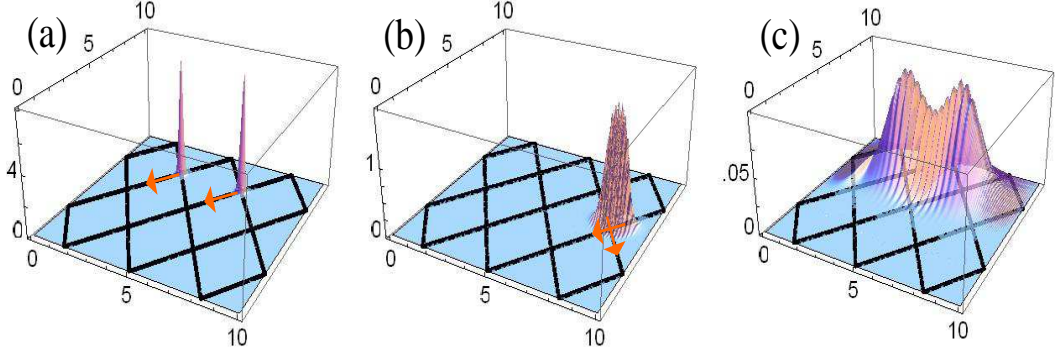}\caption{ Time evolution
of the \emph{quantum probability density} in configuration space when the
initial wavefunction is given by Eq. (\ref{4}). (a) At $t=0$ the initial
wavefunction is composed of two Gaussians centered respectively at $M$ and
$N$, each Gaussian leaving in the direction of the arrow along the classical
periodic orbit. The diagonal position density matrix elements for this initial situation corresponds to
the classical distribution portrayed in Fig. 3(a). (b) shows $\left\vert
\psi\right\vert ^{2}$ slightly before the two Gaussians cross at
$t=(11/8)T_{PO}$ (compare with Fig. 3(b)). (c) shows the situation at
$t=5T_{PO}$; the overlap of the two components of the spreading wavefunction
results in interferences relative to Fig. 3(c). }%
\label{f5}%
\end{figure}

\section{Bohmian mechanics of the square billiard}

\subsection{General remarks}

The de Broglie-Bohm theory proposes to interpret quantum phenomena in terms of
a point-like particle propagating along well-defined deterministic
trajectories in configuration space through the guidance of the wavefunction
(excellent accounts of the theory are given in Refs \cite{bohm hiley,holland}).
The initial position of the particle, and therefore its precise trajectory
cannot be known, and this is why only statistical predictions can be made;
these match the predictions of standard quantum mechanics. This is
achieved by employing the polar decomposition
\begin{equation}
\psi(\mathbf{x},t)=R_{\psi}^{1/2}(\mathbf{x},t)\exp(iS_{\psi}(\mathbf{x}%
,t)/\hbar).\label{5}%
\end{equation}
The Schr\"{o}dinger equation in terms of $R$ and $S$ yields the coupled
equations%
\begin{equation}
\frac{\partial R_{\psi}(\mathbf{x},t)}{\partial t}+\frac{1}{m}%
\mathbf{\bigtriangledown}\cdot\left(  R_{\psi}(\mathbf{x}%
,t)\mathbf{\triangledown}S_{\psi}(\mathbf{x},t)\right)  =0\label{7}%
\end{equation}
and%
\begin{equation}
\frac{\partial S_{\psi}(\mathbf{x},t)}{\partial t}+\frac
{(\mathbf{\triangledown}S_{\psi}(\mathbf{x},t))^{2}}{2m}+V(\mathbf{x}%
,t)+Q_{\psi}(\mathbf{x},t)=0,\label{6}%
\end{equation}
where $V(\mathbf{x},t)$ is the usual potential (that vanishes here except on
the billiard's boundaries) and $Q_{\psi}(\mathbf{x},t)$ is a term known as the
quantum potential given by%
\begin{equation}
Q_{\psi}(\mathbf{x},t)\equiv-\frac{\hbar^{2}}{2m}\frac{\triangledown
^{2}R_{\psi}^{1/2}}{R_{\psi}^{1/2}}.\label{9}%
\end{equation}
The momentum and the velocity of the particle are introduced via a configuration space field defined from the polar phase function through
\begin{equation}
\mathbf{p}_{\psi}(\mathbf{x},t)=m\mathbf{v}_{\psi}(\mathbf{r}%
,t)=\mathbf{\triangledown}S_{\psi}(\mathbf{x},t)\label{10}%
\end{equation}
allowing to obtain the particle's equation of motion in a pseudo-newtonian
form%
\begin{equation}
\frac{d\mathbf{p}_{\psi}}{dt}=-\triangledown(V(\mathbf{x},t)+Q_{\psi
}(\mathbf{x},t)).
\end{equation}
The defining equations of the BB theory are similar to those of classical
mechanics in the Hamilton-Jacobi formalism; compare in particular Eq.
(\ref{7}) with Eq. (\ref{z9}) and Eq. (\ref{10}) with Eq. (\ref{z8}). However
from a structural point of view, this analogy is superficial. The equations of
classical mechanics arise from the flows in phase-space obeying the canonical
equations of motion (the Hamilton equations, the principle of least action
etc.), for any choice of canonical coordinates. Any distribution can be
decomposed into elementary phase-space elements obeying these equations, i.e.
the dynamics of the distribution depends on the elementary phase-space
dynamics. In the de Broglie-Bohm theory on the other hand, the dynamics of the
particle depends on the wavefunction (this is reflected in our notation with
the indices labeled by $\psi$). The law of motion for an \emph{individual}
particle depends on the \emph{statistical distribution} $R_{\psi}$ to which
the particle belongs.\ As can be seen from Eqs. (\ref{7}) and (\ref{10}), the
dynamics of the particle is determined by the direction of the probability
flow in configuration space, which becomes the only physical representation.

\subsection{BB trajectories in the square billiard}

Although Bohmian trajectories have been computed for an incredibly wide variety of
quantum systems, and despite that fact that the particle in a 1D box is one of
the most widely used examples in any BB theory primer, very few works deal
explicitly with the determination of BB trajectories in a square billiard.
Bohm and Hiley (see Sec 8.5 of Ref. \cite{bohm hiley}) employ the square
billiard to give a general argument on the type of trajectories that can be
expected from a wavefunction obtained by combining a few eigenstates of the
Hamiltonian. This example was used as a blueprint by different authors who investigated several systems in the following
years; it was explicitly applied to the square billiard a few years later
\cite{alcantara98}. It was noted, as expected, that the type of trajectory
(regular, chaotic, localization) depended in a crucial way on how the initial
wavefunction was constructed (the choice of the participating eigenstates and
their relative weights). This was thought not to be very illuminating from the
point of view of the quantum-classical correspondence, and a further work
\cite{sales01} examined the case when a localized wavepacket was taken as the
initial wavefunction.\ Only BB trajectories for short times were computed, the
conclusion being that the Bohmian particle propagates in a classical-like way,
undergoing in particular the specular reflection on the walls. Here we give a
more complete study of BB trajectories in the framework of the
quantum-classical correspondence examined above.\ The calculations are made by integrating numerically the guidance equation (\ref{10}).
The results will be
discussed in Sec.\ 5.

\begin{figure}[tb]
\includegraphics[height=2.in,width=5.5in]{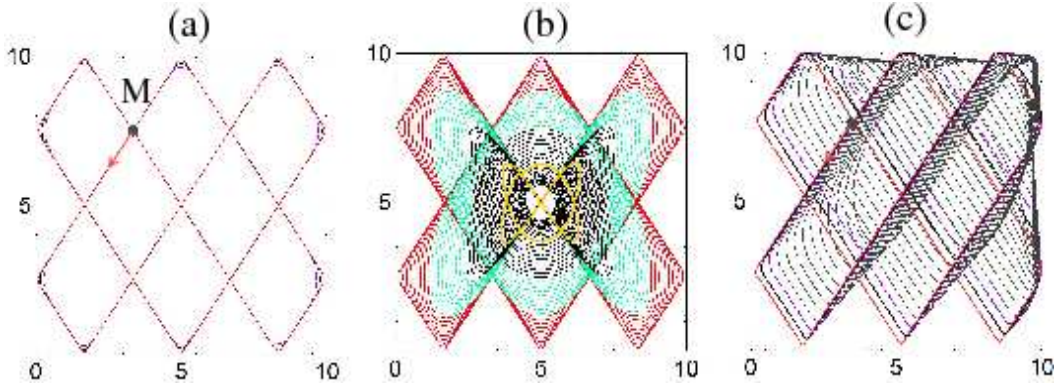}\caption{ De Broglie-Bohm
trajectories for a quantum state initially localized at $M$, whose evolution
was shown in Fig. 4. (a) Short time motion when the Bohmian particle is
initially localized at $\mathbf{x}_{i}\equiv\mathbf{x}_{M}$, the maximum of
the Gaussian. The particle follows the wavepacket, leaving $M$ in the
direction of the arrow along the red trajectory until it returns at $M$ at
time $T_{PO}$; the trajectory is quasi-periodic, and resembles the classical
periodic orbit along which the wavepacket moves. The trajectory in the
interval $T_{PO}<t<2T_{PO}$ is shown in black and in dashed purple for the
interval $2T_{PO}<t<3T_{PO}$; actually both lines are hidden behind the
previous red line except near the boundaries of the billiard, where the
Bohmian particle is reflected farther away from the boundary as time increases. (b) Same as (a)
for times up to $40T_{PO}$: the trajectory for the time interval
$0<t<10T_{PO}$ is shown in dark gray (red online), for $10T_{PO}<t<20T_{PO}$ in light dashed gray (dashed green online), and
for $20T_{PO}<t<40T_{PO}$ in black. As $t$ increases, the particle slows down;
it still follows a periodic quasi-closed trajectory, with period $T_{PO}$, but
within a zone restricted to the centre of the billiard slows down while
restricting its motion toward the center of the billiard; the thick gray (yellow online)
line represents the quasi-closed trajectory in the interval $30T_{PO}%
<t<31T_{PO}$. (c) For the same quantum state, the trajectory for a particle
with an initial position slightly off the maximum of the Gaussian
($\mathbf{x}_{i}=(x_{M}+L/80,y_{M}+L/80)$) is shown with the same colour
coding as in (a) for $0<t<3T_{PO}$ and a light gray line showing the
trajectory for longer times $3T_{PO}<t<13T_{PO}$. }%
\label{f6}%
\end{figure}

\subsubsection{Simply localized initial state}

We first compute the Bohmian trajectories in the case of Fig.\ 4 -- the
initial wavefunction is a Gaussian centered at $M$ that propagates for short
times along the classical periodic orbit shown in Fig.\ 1. The BB trajectory
when the initial position is chosen at $\mathbf{x}_{i}\equiv\mathbf{x}_{M}$
(the maximum value of the distribution) is shown for short times in
Fig.\ 6(a): the trajectory follows the wavepacket, leaving $M$ in the
direction of the arrow.\ The behaviour is nearly that of the classical
trajectory -- this is the bouncing-ball regime put in evidence in
\cite{sales01} -- and the trajectory is quasi-periodic: the BB trajectory
leaves $M$ at $t=0,$ follows the line shown in red and reaches $M$ again at
the period of the classical PO $T_{PO}$ which is also the period of the
maximum of the Gaussian wavepacket. The trajectory for $t\in\lbrack
T_{PO},2T_{PO}]$ and $t\in\lbrack2T_{PO},3T_{PO}]$ is shown in black and
dashed purple respectively -- it retraces the original path (in red) except
near the boundaries of the billiard. There, the quantum potential gets more
repulsive further from the boundaries; this is readily explained by the fact
that as the Gaussian spreads, a larger portion of the Gaussian is reflected
off the boundaries before the maximum of the Gaussian arrives. This creates an inversion of the net current in the
direction perpendicular to the wall. As a result the Bohmian particle turns around before reaching the boundary.

The same trajectory for longer times is shown in Fig.\ 6(b). The colour scheme
is the following: the trajectory for $t\in\lbrack0,10T_{PO}]$ is shown in red,
for $t\in\lbrack10T_{PO},20T_{PO}]$ in dashed green, and for $t\in
\lbrack20T_{PO},40T_{PO}]$ in black. The qualitative quasi-periodic aspect of
the trajectory disappears on a longer timescale: the particle still follows an
almost closed-orbit, though the shape of the orbit is progressively
deformed.\ The 'pseudo-period' however does not change. It takes the same
time, $T_{PO}$, for the particle to pass (very) near $M$ for the first time
after having been initially launched from $M$ than to trace an almost closed
orbit at longer times (see the thick yellow line in Fig.\ 6(b), representing
the trajectory in the interval the trajectory for $30T_{PO}<t<31T_{PO}$,
tracing a quasi-closed orbit). The particle thus slows down, restricting its
motion to a small area around the centre of the billiard.

Fig.\ 6(c) shows the Bohmian trajectory when the initial position
$\mathbf{x}_{i}$ of the particle is slightly off the maximum of the Gaussian
(the probability distribution there is about $R_{\psi}(\mathbf{x}%
_{i},0)/R_{\psi}(\mathbf{x}_{M},0)\simeq1/5$ of the maximum probability). For
short times, the Bohmian particle follows a bouncing ball regime similar to a
non-periodic classical trajectory (such as the one shown in Fig.\ 1(a)).
However after only a few pseudo-periods, as the wavefunction spreads, the
probability current drives the trajectory in the right upper quadrant of the
billiard, as the particle considerably slows down.

\subsubsection{Doubly localized initial state}

We now compute the Bohmian trajectories corresponding to the case portrayed in
Fig.\ 5: the initial wavefunction is given by Eq. (\ref{4}), with the
probability distribution being initially concentrated in the Gaussian peaks
localized at $M$ and $N$. Hence in this state, the Bohmian particle is
localized initially either near $M$ or $N$. In a given realization one of the two wavepackets is an empty wave -- it does not carry the particle but nevertheless has dynamical effects.\ Fig.\ 7(a) shows the BB trajectory
for a particle initially sitting at the maximum of the distribution
$\mathbf{x}_{i}\equiv\mathbf{x}_{M}$. The trajectory for the time interval
$0<t<T_{PO}$ is shown in dark blue in Fig. 7(a); it is also plotted in red in
Fig.\ 7(c). The trajectory leaves $M$ in the direction of the arrow, arrives
in the zone labeled $H$ in Fig.\ 7(b) with vanishing velocity; the particle
then turns around, retracing almost exactly its previous path, going back
through $M$, until it reaches the zone labeled $K$ (without crossing the
region $L$); the particle turns around with vanishing velocity at $K$ and
retraces almost its previous path until it reaches $M$ again at $t\simeq
T_{PO}$. The particle then leaves again the $M$ region in the direction of
$H$; this quasi-periodic motion is shown in Fig.\ 1(a) in light blue for
$T_{PO}<t<5T_{PO}$ and in dashed green for $5T_{PO}<t<10T_{PO} $. As $t$
increases, the trajectory turns back with increasing distance from the regions
around $H,$ $L$ and $K$, and the mean velocity decreases.

Fig.\ 7(b) shows the BB trajectory for a particle initially sitting at the
maximum of the distribution $\mathbf{x}_{i}\equiv\mathbf{x}_{N}$, with the
same colour scheme employed in Fig.\ 7(a). The deformation of the
quasi-periodic trajectory as $t$ increases is identical to the one seen in
Fig.\ 7(a), except that the particle occupies a different area of the square
billiard. The trajectory for $0<t<T_{PO}$ is also shown in dark blue (dark
grey) in Fig.\ 7(c): the particle initially leaves $N$ in the direction of the
arrow, turns back in the $H$ region, going back through $N$ until it reaches
$K$ (without crossing the $L\ $region), at which point it turns back and
reaches $N$ again at $t\simeq T_{PO}$. Note that \emph{taken together}, the
two trajectories for Bohmian particles initially placed at $M$ and $N$ trace,
in the interval $0<t<T_{PO}$, the shape of the classical periodic orbit going
through $M$ and $N$ (compare Fig.\ 7(c) with Figs.\ 6(a) and 1(a)). This is to
be expected, since for short times, the distribution does not spread
significantly and each wavepacket initially centered at $M$ and $N $ moves by
following the periodic orbit. Of course this is not the case for longer times:
Fig.\ 7(d) shows the same Bohmian trajectories initially launched from $N$ and
$N$ for times $10T_{PO}<t<11T_{PO}$ (compare with Fig.\ 6(b)).

\begin{figure}[tb]
\includegraphics[height=4.2in,width=4.7in]{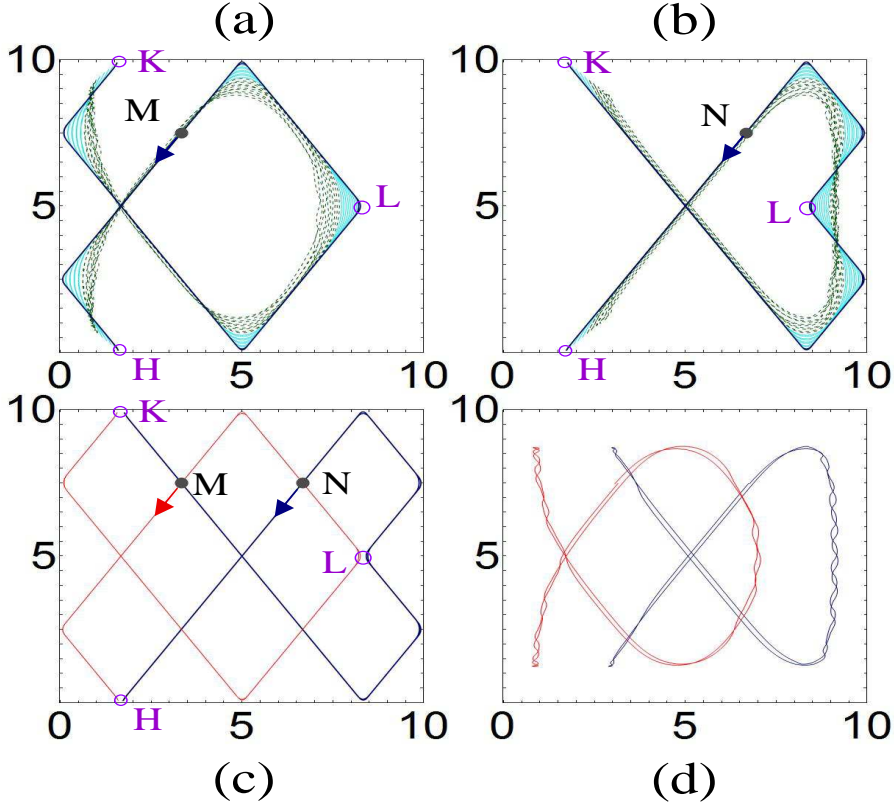}\caption{ De Broglie-Bohm
trajectories for the quantum state given by Eq. (\ref{4}), initially localized
at $M$ and $N$ whose evolution was shown in Fig. 5. (a) Trajectory obtained
when the Bohmian particle is initially localized at $\mathbf{x}_{i}%
\equiv\mathbf{x}_{M}$. The particle leaves with the wavepacket in the
direction of arrow, turns around at $H$, then $K$, and arrives at $M$ in a
time $T_{PO}$ (dark line, dark blue online). The motion in the time intervals
$T_{PO}<t<5T_{PO}$ and $5T_{PO}<t<10T_{PO}$ is shown in light gray (light blue online) and dashed
dark gray (dashed green online) respectively. (b) Same as (a) but when the Bohmian particle is
initially localized at $\mathbf{x}_{i}\equiv\mathbf{x}_{N}$. (c) Bohmian
trajectories for $0<t<T_{PO}$ taken from (a) and (b) plotted together; the
overall shape is very similar to that of the classical periodic orbit of Fig.
1 going through $M$ and $N$. (d) Same as (c) for $10T_{PO}<t<11T_{PO}$. }%
\label{f7}%
\end{figure}

The main feature visible for the BB trajectories in the doubly localized
initial state is the reflection that the particle undergoes at $H$, $L$ and
$K$. This behaviour has no classical counterpart: it is the result of the
interference of the two Gaussian amplitudes at these points. Indeed the center
of the two Gaussians cross at these points; Fig.\ 5(c) illustrates the case of
a constructive interference at $L$. Here the two wavepackets have opposite
components along the $x$ axis, so the net probability density current along
$x$ decreases as the two Gaussians start to interfere. The particle, moving
before the crossing point along with the localized wavepacket, slows down,
whereas the Gaussians keep moving with constant group velocity. Ultimately the
net current in the $x$ direction reverses before the particle reaches the
crossing point, thereby giving rise to the avoided crossing type of behaviour
at $L$. When the net current is reversed in both the $x$ and $y$ directions,
the Bohmian particle's motion is reversed; this is what happens in the $H$ and
$K$ regions.\ In this respect, it may be noted that the empty wavepacket (the
Gaussian that, before the crossing, does not carry the particle) produces the
same dynamical effect than the reflection on the billiard's
boundaries.\ Indeed, at $H$ and $K$ the reversal of the probability current
along $x$ is due to the empty wavepacket, whereas the reversal in the $y$
direction is produced by the reflection of the Gaussian wavepacket on the
boundary walls.

\section{Discussion}

\subsection{The quantum-classical correspondence}

The term 'quantum-classical correspondence', as employed in this paper,
reflects the use that is usually made in works undertaken in the field of
'quantum chaos'. It is indeed well-known that when the classical action
appearing in the path integral expression for the propagator $K(\mathbf{x}%
_{0},\mathbf{x},t)$ is much larger than $\hbar$, then $K(\mathbf{x}%
_{0},\mathbf{x},t)$ can be approximately written in terms of the sole
classical paths relating $\mathbf{x}_{0}$ to $\mathbf{x}$. This allows to
understand the properties of quantum systems (such as the distribution of the
energy levels, scars in the wavefunctions, recurrences in the time-dependent
behaviour of the system) in terms of the properties of the analog classical
system \cite{haake,gutz}. Such a correspondence is particularly useful in the
studies of nonseparable quantum systems, for which quantum computations are
either nonfeasible or yield uninterpretable results.

In the square billiard, the quantum-classical correspondence is extremely
simple: the semiclassical form of the propagator (\ref{z21}) is exact meaning
that each point of the quantum wavefunction propagates along a classical
trajectory. This explains the correspondence between the dynamics of the
classical distributions (Figs.\ 2 and 3) and those of the quantum
distributions (Figs. 4 and 5). The dynamical evolution in the classical and
quantum cases is essentially identical, the important difference being of
course that classically the probability \emph{distribution} evolves by
following the classical trajectories, whereas quantum mechanically the
\emph{wavefunction} evolves by following the classical trajectories, the
distribution resulting from the interference of different bits of the
wavefunction each carried by a classical trajectory.

Hence the quantum-classical correspondence is only dynamical -- it remains
silent on the ontological aspects.\ The classical ensemble is a statistical
distribution of point-like particles in phase-space, giving the probability
distribution in configuration space as a function of time for a fixed initial
distribution. The quantum ensemble appears as the intensity of a field (here
presented in configuration space). Although the field moves by following the
classical motion, it cannot be associated with a distribution of point-like
particles -- the field interferes and only its intensity represents the
statistical distribution.

\subsection{Properties of Bohmian trajectories}

From Eqs. (\ref{7}) and (\ref{10}), it is straightforward to establish
\cite{holland} that the de Broglie-Bohm trajectories follow the streamlines
of the probability flow, the velocity of the Bohmian particle depending on
the local current density $\mathbf{j}_{\psi}$ through%
\begin{equation}
\mathbf{j}_{\psi}(\mathbf{x},t)=R_{\psi}(\mathbf{x},t)\mathbf{v}_{\psi
}(\mathbf{x},t).
\end{equation}
Hence if the wavefunction is a localized wavepacket, the particle will be
restricted to move with the wavepacket; in a coarse-grained sense, the
particle's motion follows that of the wavepacket, irrespective of its precise
initial position. This is particularly clear in Fig. 6(a): the wavepacket
initially sharply localized at $M$ follows the classical periodic orbit of
Fig.\ 1(a). The BB trajectory therefore also follows the classical trajectory,
except near the boundary of the billiard where the part of the wavepacket that
is traveling ahead of the Bohmian particle's position is reflected, thereby
reversing the current density and preventing the particle from reaching the boundary.

As the wavepacket spreads over substantial portions of the billiard, each
Bohmian trajectory will be sensitive to very fine details of the probability
flow, implying a crucial dependence of the trajectory on the initial
distribution and the initial position. In Fig.\ 6(b), when the initial
position lies on the top of the Gaussian, the ensuing trajectory is symmetric
around the center of the billiard, whereas when the initial position is
slightly off the center of the Gaussian (Fig. 6(c)) the fixed point is in the
upper right-hand corner. Given that locally the probability flows have no
relation to the classical dynamics of the billiard, the BB trajectories will
be unrelated to the classical ones.

The example portrayed in Fig.\ 7, corresponding to a doubly localized initial
state, shows that even for short times and localized wavepackets, BB
trajectories can be markedly different from the classical ones. This is well
illustrated in Fig.\ 7(c): although each of the localized wavepackets launched
from $M$ and $N$ follows the classical periodic orbit of Fig.\ 1(a), when
these wavepackets cross the net current density can locally vanish and reverse
its course. This is why taken globally, the two Bohmian trajectories are able
to recover the periodic orbit, although an individual Bohmian
particle travels only by following one of the two trajectories depending on
its initial position. This is an essential consequence if the dynamics predicted by
the de Broglie-Bohm theory is taken as a realist account of quantum phenomena.

\subsection{Bohmian mechanics and the classical limit}

It is well-known that typical Bohmian trajectories are not classical.\ As
such, this feature is not necessarily a problem, provided one can account
unambiguously for the emergence of classical mechanics. However achieving the
classical limit turns out to be an intricate problem (see Ch.\ 6 of
\cite{holland}). It is generally admitted
\cite{bohm hiley,holland96,appleby,bowman,lan} that BB trajectories in closed systems being
generically non-classical, in order to emerge, classical mechanics
calls for a special class of states combined with environmental interactions.
More precisely, following an initial suggestion made by Bohm and Hiley
\cite{bohm hiley}, it has been argued \cite{appleby,bowman} that Bohmian
mechanics in the classical limit must involve a mechanism yielding a localized
and non-spreading packet behaving quasi-classically. This mechanism can only
arise when the closed system interacts with an environment inducing decoherence.

Relying on decoherence to recover classical trajectories from non-classical
ones raises a number of problems. First, as we have discussed elsewhere
\cite{amvn}, this argument appears as somewhat specious when applied to
semiclassical systems: in such closed quantum systems, the wavefunction as
well as several observable properties display the fingerprints of the
underlying classical dynamics.\ It is then difficult to explain why the
supposedly real motion followed by the Bohmian particle is non-classical (and
only takes a classical-like form when interacting with an environment)
although the wavefunction of the closed system evolves dynamically in a classical manner and
displays classical morphological features. In this respect, the square billiard is
a simple semiclassical system, for which, as shown above, the
quantum-classical correspondence is straightforward; several quantities not
discussed in the present work (like recurrences in the autocorrelation
function) can also be given a semiclassical explanation in terms of the
large-scale structures determined by the underlying classical dynamics.
As already mentioned (see also \cite{pla05,pla07}), the de Broglie-Bohm
account of such features involves non-classical trajectories following the
streamlines of the flow. Only globally are the classical structures recovered
on a statistical basis, like the classical periodic orbit obtained in
Fig.\ 7(c) by combining two individual non-classical Bohmian trajectories.

The second point concerns the specific conditions under which decoherence will turn Bohmian into classical trajectories. Only tentative models were given
in \cite{bohm hiley,appleby,bowman} all of them relying on obtaining non-spreading wave-packets
behaving quasi-classically.  But as a general rule decoherence converts a
pure state into a (in practice) mixed state without necessarily implying non-spreading probability
distributions. Moreover, the existence of models for which
decoherence successfully does the job of recovering classicality would not constitute a specific asset for the de Broglie-Bohm interpretation: any other
interpretation of quantum phenomena (such as the multiple-universe
interpretation) that can dispose of Leggett's 'logical fallacy' \cite{leggett
02} concerning the reinterpretation of the wavefunction post-decoherence is as
effective as Bohmian mechanics in achieving classical motion. Indeed, according to this account, the
existence of quantum trajectories is not needed to explain the
emergence of classicality.

More fundamentally, in line with the first remark it can be objected
\cite{ballentine} whether requiring localized and non-spreading mixtures is
the proper way to achieve the classical limit. Indeed we have seen that
classical distributions do spread; and in the case of the square billiard they
spread exactly like the amplitude of the quantum distribution (i.e. the
smoothed non-oscillating quantum distribution). Classically, the spreading is the result of
the Liouville diffusion of the statistical distribution. Since the
quantum-mechanical statistical distribution depends on the
wavefunction, there is no reason to constrain the wavefunction to avoid
spreading, unless one phenomenologically associates a particle with a
localized non-spreading wavepacket in an ad-hoc way (by stating for example that what appears classically as particles is nothing but the quantum mechanics of localized non-interfering wavepackets).\ Clearly, such an
association would be inconsistent with the de Broglie Bohm interpretation, where the existence of point-like particles is postulated, and
where the statistical role of the wavefunction modulus -- the term that spreads --
is particularly emphasized. But then, since spreading necessarily brings in
interferences (e.g. when the wavepacket hits the billiard's boundary), it becomes
impossible to recover classical motion if one maintains that the spreading
wavefunction's amplitude represents the statistical distribution of a particle
moving along a given streamline of the flow. This conclusion was already put
forward in different terms by Holland \cite{holland96} who noted that neither
pure nor mixed states allowed to generate classical motion from the ensuing
probability flow.

\section{Conclusion}

In this work, we have investigated Bohmian trajectories in a square billiard
and contrasted them with quantum wavepacket dynamics and with the
trajectories of the classical square billiard. As expected from a path
integral approach, the quantum square billiard displays dynamics abiding by
the quantum-classical correspondence -- each bit of the spreading quantum
wavefunction propagates along a classical trajectory. On the other hand,
individual Bohmian trajectories were shown to be generically highly
non-classical, although statistically the underlying classical large scale
structures are recovered as expected. We have also argued that the inclusion
of decoherence is unlikely to allow de Broglie-Bohm dynamics to recover classicality. This conflict between the dynamical continuity involving the classical propagation of the wavefunction and the persistence of non-classical, typically quantum features of the probability distribution on smaller scales is not limited to the de Broglie-Bohm interpretation -- it is relevant to the study of the quantum-classical transition irrespective of any particular interpretation. However, this conflict does take an acute form for Bohmian mechanics because the ontological claims made by this interpretation involve a continuity with the ontology of classical mechanics.

\end{document}